\documentstyle[eqsecnum,epsf,aps,twocolumn]{revtex}

\begin{document}
\draft

\title{Dynamics of a heavy particle in a Luttinger liquid}
\author{A.H. Castro Neto$^{1,2}$ and Matthew P.A. Fisher$^1$}
\address{$^1$
Institute for Theoretical Physics, University of California\\
Santa Barbara, CA 93106-4030}
\address{$^2$
Department of Physics, University of California\\
Riverside, CA 92521}

\date{\today}
\maketitle

\begin{abstract}
We study the dynamics of a heavy particle of mass $M$ moving in a
one-dimensional repulsively interacting Fermi gas.
The Fermi gas is described using
the Luttinger model and bosonization.
By transforming to a frame co-moving with the heavy particle,
we map the model onto a generalized ``quantum impurity problem".
A renormalization group calculation reveals a crossover from strong to weak
coupling upon scaling down in temperature.
Above the crossover temperature scale
$T^* = (m/M)E_F$,
the particle's mobility, $\mu$, is found
to be (roughly) temperature independent and proportional
to the
dimensionless conductance, $g$, characterizing the 1d Luttinger liquid.
Here $m (<<M)$ is the fermion mass, and $E_F$ the Fermi energy.
Below $T^*$, in the weak coupling regime,  the mobility grows and
diverges as $\mu(T) \sim T^{-4}$ in the $T \to 0$ limit.
\end{abstract}

\bigskip


\narrowtext

\section{Introduction}
The quantum dynamics of a heavy particle moving through a fluid
has been of longstanding interest.
Most of the effort has focussed on
three dimensional quantum fluids,
either Fermi liquids such as
${}^3 He$ or superfluids such as ${}^4 He$ \cite{nikolai}.
Recently, there has been a resurgence of interest
in non-conventional quantum liquids.
A paradigm is the Luttinger model\cite{lut}, which
describes a one-dimensional interacting Fermi gas.

In this paper we study in detail the dynamics
of a single heavy particle moving through a 1d
Luttinger liquid.  Of interest is the temperature
dependence of the heavy particle's mobility.
Our motivation is two-fold.
Firstly, since the excitations in a 1d Luttinger liquid are
profoundly different
than in a Fermi liquid, one might anticipate
that the dynamics of an immersed heavy particle would
likewise be qualitatively modified.  Secondly,
powerful non-perturbative methods in 1d, such as bosonization,
might be fruitfully employed to
analyze the dynamics of a strongly coupled heavy particle.

Our main results are as follows.
After introducing the model in Section II,
we transform to a frame of reference co-moving
with the heavy particle in Section III.
In this frame, the heavy particle sits at the origin.
In the limit that $M \to \infty$ the model
then becomes equivalent to a Luttinger liquid
scattering off a static localized impurity.
This problem has been analyzed in great detail recently,
and is now well understood \cite{kf}.
In the zero temperature limit, the impurity effectively ``breaks"
the Luttinger liquid into two semi-infinite de-coupled
pieces.  Fermions incident on the
impurity are completely reflected.  To analyze
the case with finite mass $M$,
a natural starting point is thus a limit in which
the amplitude $t$ for incident fermions to
tunnel through the heavy particle is set to zero.
Provided $t=0$, the mobility can be computed for arbitrary $M$,
and one finds a temperature independent value,
$\mu = \pi g /(2 \hbar k_F^2)$.  At low temperatures, though,
this limit is unstable to non-zero tunnelling, $t$.
A renormalization group calculation reveals a crossover to
a regime where the fermions are transmitted readily through
the heavy particle.  This leads to a decoupling between
the dynamics of the heavy particle and the Fermi sea.
At zero temperature, the only effect
of the Fermi sea is to renormalize the mass of the
heavy particle - the mobility is infinite.

In Section IV we use a weak coupling perturbative approach
to calculate the temperature dependence of the mobility
in the $T \to 0$ limit.  The dominant scattering process involves
four fermions, absorbing and then re-emitting
a pair, one right and one left moving.
This process changes the momentum and energy of the heavy particle,
and is shown to lead to a low temperature mobility
that diverges as $\mu(T) \sim T^{-4}$.

\section{The Model}

The
Hamiltonian which describes the motion of a heavy particle
coupled to a 1d interacting Fermi gas can be written
as $H = H_0 + H_{LL} + H_{int}$.
Here $H_0$ describes the free particle of mass $M$:
\begin{equation}
H_0 =\frac{P^2}{2 M} ,
\end{equation}
with momentum $P$ and position $X$.
$H_{LL}$ is the Hamiltonian for N-interacting fermions,
which in first quantized notation is,
\begin{equation}
H_{LL} =\sum_{i=1}^{N}\frac{p_i^2}{2 m} +\sum_{i,j} V(x_i-x_j)
\end{equation}
where $x_i$ and
$p_i$ denote coordinate and momentum of the $i^{th}$ particle.
The interaction between the heavy particle and fermions is assumed to take the
form,
\begin{equation}
H_{int} = \sum_{i=1}^{N} U(x_i-X) .
\end{equation}
For simplicity we assume that
$U(x)$ is repulsive and short-ranged.

It is useful also to have a 2nd quantized formulation.
We denote as $\psi(x)$ the fermionic field operator describing
the interacting Fermi gas.  In the absence of interactions,
the ground state consists of a filled Fermi sea, with
Fermi
momentum $k_F$.  As usual, we decompose the field into a sum of right
and left movers:
\begin{equation}
\psi(x) = \psi_R(x) e^{ik_Fx} + \psi_L(x) e^{-ik_Fx}
\end{equation}
where $\psi_{R/L}$ are supposed to be slowly varying.
It will also be useful
to bosonize the interacting electron gas,
by expressing
\begin{equation}
\psi_{R/L}(x) = \sqrt {k_F} e^{i \sqrt {\pi} (\phi \pm \theta)} ,
\end{equation}
where $\phi$ and $\theta$ are canonically conjugate fields satisfying
$[\theta(x), \partial_x \phi(x')] = \delta(x-x')$.
The appropriate Luttinger liquid Hamiltonian
takes the from:
\begin{equation}
H_{LL} = \frac{v}{2} \int_x [ \frac{1}{g} (\partial_x \theta)^2 +
g (\partial_x \phi)^2 ] .
\end{equation}
Here $g$ is the dimensionless conductance, which is less than one
for a repulsively interacting Fermi gas, equals one
for free fermions,
and is greater than one with attractive interactions.  The Fermi velocity
$v$ is also renormalized
by interactions, and will differ from the free fermion value,
$k_F/m$.

The right and left moving electron densities, $N_{R} =
\psi^\dagger_{R} \psi_{R}$
and $N_{L} =
\psi^\dagger_{L} \psi_{L}$ have simple bosonic representations,
\begin{equation}
N_R + N_L = (1/\sqrt {\pi}) \partial_x \theta
\end{equation}
and
\begin{equation}
N_R - N_L = (1/\sqrt {\pi}) \partial_x \phi  .
\end{equation}

\section{Description in Frame co-moving with particle}

To transform the equations of motion into a frame co-moving with the heavy
particle, one can use the
unitary transformation,
\begin{equation}
{\cal U} = e^{i \sum_{i=1}^{N} p_i X}.
\end{equation}
This transformation has been previously used\cite{zotos}
in a similar context, but in the special case
where $M=m$ .
Under this transformation, the coordinates and momenta transform as:
\begin{eqnarray}
x_i &\to& x_i + X ,
\nonumber
\\
p_i &\to& p_i  ,
\nonumber
\\
X &\to& X  ,
\nonumber
\\
P &\to& P - \sum_{i=1}^{N} p_i.
\end{eqnarray}
The transformed Hamiltonian becomes,
\begin{equation}
H \to H = \frac{1}{2M} [ P- \sum_{i=1}^{N} p_i ]^2 +
H_{imp},
\end{equation}
with
\begin{equation}
H_{imp} = H_{LL} + \sum_{i=1}^{N} U(x_i)  .
\end{equation}
Notice that when $M \to \infty$
the full Hamiltonian reduces to $H_{imp}$,
which  describes a Fermi gas interacting with a static
potential, $U(x)$, centered at the origin.  This quantum
impurity problem has recently been analyzed in great detail\cite{kf}.
However,
when $M$ is finite the heavy particle can move, and exchange energy
with the Fermi sea.  Notice that the heavy particle is
coupled to the fermions via a
minimal coupling, where the ``gauge" field is the
total momentum in the Fermi sea \cite{nos,pre,prb,prl}.

The transformed Hamiltonian can be expressed
directly in 2nd quantization using the fermion field
operators (2.4).  These fields can then be bosonized.
It is convenient to use a path integral representation,
since the Lagrangian is linear in the ``gauge" field.
The euclidean action for the free Luttinger liquid
which corresponds to (2.6) can be expressed as,
\begin{equation}
S_{LL} = \frac{g}{2v} \int_{x,\tau}
[v^2 (\partial_x \phi)^2 + (\partial_\tau \phi)^2 ]  .
\end{equation}
The total momentum of the fermions can also be easily bosonized,
\begin{equation}
\sum_{i=1}^{N} p_i \to k_F \int_x (N_R-N_L) = \frac {k_F}{\sqrt{\pi}}
\int_x \partial_x \phi  .
\end{equation}
which enables the total action for the heavy particle plus
Luttinger liquid to be written,
\begin{equation}
S= \frac{M}{2} \int_\tau \dot{X}^2(\tau) + \frac{i k_F}{\sqrt{\pi}}
\int_{x,\tau} \dot{X}(\tau)
\partial_x \phi(x,\tau) + S_{imp}.
\end{equation}

To analyze
the dynamics in the transformed frame, it is convenient
to first consider a strong coupling limit ($U \to \infty$).
In this limit, the fermions cannot pass through the heavy particle,
and the Luttinger liquid is divided into two de-coupled
regions
on either side of the particle.
Perturbations away from this limit
can be included by allowing for tunnelling of fermions
from one side to the other, with a small amplitude $t$.
This process can be expressed in terms of the bosonic fields as \cite{kf},
\begin{equation}
S_T = - t \int_\tau \cos \sqrt{\pi} \left(
\phi(0^+,\tau) - \phi(0^-,\tau)\right)  .
\end{equation}
As we shall see, in the limit $t=0$ the heavy particle's dynamics
can be obtained exactly.  A perturbative analysis for small $t$
is then possible.

To this end, we follow
ref.\cite{kf} and integrate out the bosonic field $\phi(x)$,
except at $x=0$ - that is at the position of the heavy particle.
In terms of the phase difference across the heavy particle,
\begin{equation}
\Phi(\tau) = {1 \over 2} [\phi(0^{+},\tau)-\phi(0^{-},\tau)],
\end{equation}
the action becomes $S=S_0 + S_T$ with
\begin{equation}
S_0 = \frac{1}{\beta}
\sum_n \left(\frac{M \omega_n^2}{2} |X_n|^2 +
\frac{2k_F \omega_n}{\sqrt{\pi}} X_n \Phi_{-n}
+ g |\omega_n| |\Phi_n|^2 \right) ,
\end{equation}
\begin{equation}
S_T = - t \int_\tau \cos(2 \sqrt{\pi} \Phi(\tau)) .
\end{equation}
In (3.10) the summation is over
Matsubara frequencies $\omega_n = 2\pi n/ \beta$,
with $\beta$ the inverse temperature.

In the limit of zero tunnelling ($t=0$), the action is quadratic.
One can then integrate over the field $\Phi(\tau)$,
to obtain a simple action for the
dynamics of the heavy particle,
\begin{equation}
S_{X} = \frac{1}{\beta} \sum_n \left(\frac{M \omega_n^2}{2}
+ \frac{ k_F^2 |\omega_n|}{\pi g}
\right)|X_n|^2.
\end{equation}
This action is of the Caldeira-Leggett form,
and describes a particle undergoing
Brownian motion in a viscous environment with friction coefficient,
$\eta = \frac{2 k_F^2}{\pi g}$\cite{caldeira,aoah}.
The particle's mobility
can be obtained from the Kubo formula \cite{kubo},
\begin{equation}
\mu(\omega) = \frac{1}{\omega_n} P(\omega_n) |_{\omega_n \to i\omega +
\epsilon},
\end{equation}
\begin{equation}
P(\omega_n) =
\int_\tau e^{-i \omega \tau} <\dot{X}(\tau)
\dot{X}(0)> = <|X_n|^2> .
\end{equation}
For the quadratic action (3.12) this gives a dc mobility,
\begin{equation}
\mu = \frac{\pi g}{2 \hbar k_F^2} \, \, ,
\end{equation}
which is independent of temperature and proportional to the
Luttinger liquid conductance $g$.
In this limit ($t=0$), the particle is heavily damped
by the fermions, even at zero temperature.
The damping is heavy because the fermions cannot
pass through the heavy particle,
so motion is only possible by ``pushing" the fermions out of the way.

Consider now perturbing about this limit, for small tunnelling $t$.
We first integrate over $X(\tau)$, to obtain an action which depends only
on the bosonic field $\Phi$:
\begin{equation}
S_{\Phi} = \frac{1}{\beta} \sum_n \left(\frac{2 k_F^2}{\pi M}
+ g |\omega_n|\right) |\Phi_n|^2
+ S_T  .
\end{equation}
Notice that the phase mode has a mass term, due to
the motion of the heavy particle.  In the static limit
($M \rightarrow \infty$) this mass term vanishes,
and the action reduces to that for a Luttinger liquid with
impurity.
Consider now a renormalization group (RG) transformation
which consists of integrating over modes $\Phi(\omega)$,
for frequencies between $\Lambda/b$ and $\Lambda$,
and then rescaling $\omega \to \omega' = \omega/b$.
Here $\Lambda \sim E_F$ is a high frequency cutoff,
and $b=e^{dl}$ is a rescaling factor.
This transformation leaves the coefficient $g$ invariant,
whereas $M$ decreases as,
\begin{equation}
\frac{d M}{d l} = - M  .
\end{equation}
The RG flows for $t$ depend on whether the mass
for the phase mode is larger or smaller than the cutoff
$\Lambda$.  For
$M >> \frac{k_F^2}{\Lambda}$ the lowest order RG flow equation is
\begin{equation}
\frac{d t}{d l}  = \left(1-\frac{1}{g}\right) t
\end{equation}
whereas for $M << \frac{k_F^2}{\Lambda}$ one has,
\begin{equation}
\frac{d t}{d l} = t .
\end{equation}
At finite temperatures, these RG flows will be cut-off
at a scale $b \sim \Lambda/T$.

Since the cutoff energy scale is essentially the Fermi energy,
$\Lambda \sim k_F^2/m$, the crossover between the two
flows occurs when $M(l)\sim m$.
If the (bare) particle mass is very large, $M>>m$,
the scaling of $t$ will be determined by (3.18) over
a large range of temperatures, between $E_F$
and a crossover scale $T^* \sim (m/M)E_F$.
In this temperature range, for a repulsively interacting
Luttinger liquid ($g<1$), the tunnelling rate
will scale towards zero.  The mobility of the heavy particle
should then be roughly independent of temperature, given by (3.15).
However, at temperatures below $T^*$, (3.19) indicates that
the tunnelling rate $t$ starts increasing.
As $T \rightarrow 0$ the tunnelling rate becomes large,
and the perturbative expansion breaks down.

Evidently, in the low temperature limit the fermions can tunnel easily
through the heavy particle.  One anticipates that as $T \rightarrow 0$
the heavy particle becomes transparent,
and it's dynamics decouples from the fermions.

At very low temperatures when $t$ grows large,
fluctuations in the phase $\Phi$
are greatly suppressed by the $S_T$ term in (3.16).
In this limit it is a good approximation
to expand the cosine in (3.11) for small argument:
\begin{equation}
-t \cos(2 \sqrt{\pi} \Phi) \to -t + 2 \pi t \, \Phi^2.
\end{equation}
This explicitly breaks the $2\pi$ phase invariance of the
action.  This symmetry breaking presumably occurs
spontaneously at $T=0$, but would be
restored at non-zero $T$.
This approximation is thus only expected to be strictly valid {\it at} $T=0$.
Since each $2 \pi$ phase-slip process represents an event in which
a fermion backscatters off the heavy particle, these events
are completely suppressed at $T=0$.

After expanding the cosine term the full action is quadratic,
\begin{equation}
S_{\Phi} = \frac{1}{\beta} \sum_n
\left(\frac{2  k_F^2}{\pi M} +2 \pi t + g |\omega_n|\right) |\Phi_n|^2  .
\end{equation}
The mobility can then be calculated using (3.14).  To this end
we introduce a source term,
$S_s = i \int d\tau \dot{X}(\tau) J(\tau)$,
which enables us to express $P(\omega_n)$ as a correlation function
over the phase field:
\begin{equation}
P(\omega_n) = \frac{1}{M} \left(1- \frac{2  k_F^2}{\pi M}
<|\Phi_n|^2>\right).
\end{equation}
This can be evaluated using (3.21) and one finds,
\begin{equation}
P(\omega_n) = \frac{1}{M} \left(\frac{2 \pi t + g|\omega_n|}{
(2 k_F^2/\pi M) + 2 \pi t + g |\omega_n|}\right) .
\end{equation}
When $t<<\omega_n <<(m/M)E_F$, this reduces to our previous result (3.15).
However, in the low frequency limit, $\omega_n << t, (m/M)E_F$,
it gives a diverging ac mobility:
\begin{equation}
\mu(\omega) = \frac{1}{i\omega M_{eff}} ,
\end{equation}
\begin{equation}
M_{eff} = M (1+ \frac{2(m/M)E_F}{\pi^2 t} )  .
\end{equation}
This describes ballistic motion of the heavy particle
with an effective mass, $M_{eff}$.
This result is
valid only {\it at} $T=0$.
At non-zero but small temperatures, $T<<(m/M)E_F$,
one expects a finite mobility.
As will be confirmed in Section IV, the d.c. mobility
indeed diverges as $T \to 0$.

The above results suggest a rich temperature
dependence for the mobility for $g<1$.
Between the Fermi temperature and a crossover temperature,
$T^* \sim (m/M) E_F$, the mobility is roughly temperature
independent and given by (3.15).
Below $T^*$, the mobility starts increasing with cooling,
and diverges in the zero temperature limit.  Physically,
below $T^*$ the heavy particle becomes ``transparent"
to the fermions.  The dynamics
of the heavy particle decouples from the Fermi sea.
In the next Section, we employ a weak coupling
perturbative approach to calculate the functional form of $\mu(T)$
as $T \to 0$.

\section{Weak coupling perturbation theory}

Since the heavy
particle tends to decouple from the Fermi sea
as $T \rightarrow 0$, a weak coupling approach
should be appropriate at low temperatures.
In this Section we use perturbation theory
in the coupling between particle and Fermi sea, to
extract the temperature dependence of the mobility
as $T \rightarrow 0$.

It is convenient to employ
a 2nd quantized description for the heavy particle,
denoting as $c^{\dagger}(x)$ and $c(x)$ the creation and destruction
operators.  Since we are only interested
in a {\it single} particle, $\int_x c^\dagger c =1$.
The free Hamiltonian (2.1) is,
\begin{equation}
H_0 = \sum_k \epsilon_k c^\dagger_k c_k
\end{equation}
with dispersion $\epsilon_k = k^2/2M$.
The interaction Hamiltonian (2.3) becomes
\begin{equation}
H_{int} = U_0 \int dx c^\dagger (x) c(x) N(x) ,
\end{equation}
where $N(x) = \psi^\dagger(x) \psi(x)$ is the fermion density.
Here we have replaced the short-ranged interaction by a delta-function:
$U(x) \rightarrow U_0 \delta(x)$.
It is important to distinguish between small momentum transfer
processes, and processes which scatter the fermions by $2k_F$.
Using the decomposition (2.4) one can
express, $N= N_0 + N_{2k_F}$, where
\begin{equation}
N_0(x) = \psi^\dagger_R \psi_R + \psi^\dagger_L \psi_L = N_R + N_L  ,
\end{equation}
involves small momentum transfer,
and
\begin{equation}
N_{2k_F}(x) = \psi^\dagger_R \psi_L e^{i2k_Fx} + h.c. ,
\end{equation}
denotes the large momentum contributions.
The two corresponding terms generated from the interaction Hamiltonian will be
denoted
$H_{int,0}$ and $H_{int,2k_F}$, respectively.

Consider first computing the scattering rate for the heavy particle
using Fermi's Golden rule, where the perturbing Hamiltonian is
$H_{int,0}$.  Since the fermion density at small momentum
transfer is simply, $N_0 = (1/\sqrt {\pi}) \partial_x \theta$,
the interaction
Hamiltonian, $H_{int,0}$, takes the form of an ``electron-phonon"
interaction.
It is thus useful to introduce ``phonon" creation and destruction operators,
which create and destroy the Harmonic Luttinger liquid
excitations.
To this end, we expand the boson field as
\begin{equation}
\theta(x) = \frac{1}{\sqrt{L}}\sum_k
\theta_k e^{ikx}
\end{equation}
and introduce boson operators:
\begin{equation}
b_k = (1/\sqrt{2g |k|}) (|k|\theta_k + ig\Pi_k )  ,
\end{equation}
where $\Pi_k$ denote Fourier modes of the conjugate momentum,
$\Pi(x) = \partial_x \phi(x)$.
The operators $b_k$ satisfy canonical Bose commutation relations.
The Luttinger liquid Hamiltonian can be expressed as,
\begin{equation}
H_{LL} = \sum_k \omega_k b^\dagger_k b_k  ,
\end{equation}
with dispersion, $\omega_k = v|k|$.
Finally, the small momentum interaction takes the form:
\begin{equation}
H_{int,0} = \frac{\sqrt{v} U_0}{\sqrt{\pi L}} \sum_{k,q}
(iq/\sqrt{\omega_q}) c^\dagger_{k+q}c_k
(b_q + b^\dagger_{-q})  .
\end{equation}

The rate to scatter the heavy particle
from an initial state with momentum $k$ to
a final state $k'=k+q$, with absorption or emission
of a single phonon,
can now be readily obtained using Fermi's Golden rule.
After summing over all possible
phonon modes, assuming they are in equilibrium
at temperature $T$,  the rate is found to be,
\begin{equation}
\Gamma_{k \rightarrow k+q} = \frac{U_0^2}{vL}
\sum_{\pm} \omega_q (2n_q +1 \pm 1)
\delta(\epsilon_{k+q} - \epsilon_k \pm \omega_q)  .
\end{equation}
Here $n_q = (exp(\beta \omega_q) -1)^{-1}$ is the Bose distribution function.

These processes are severely restricted by energy and momentum conservation.
For example, for zero initial momentum, $k=0$, the
above delta functions vanish unless, $\epsilon_q = \omega_q$,
or $q=2Mv_F$.  But at this momentum, the heavy particle
has energy, $\epsilon_{q=2Mk_F} = (4M/m)E_F$.
These processes will thus freeze out exponentially fast for
temperatures below this energy scale.  If these were the only processes
present, the mobility would diverge exponentially in the $T \rightarrow
0$ limit.  But other processes will dominate at low temperatures,
as we now discuss.

Consider next the $2k_F$ scattering contribution,
\begin{equation}
H_{int,2k_F} = U_0 \int_x c^\dagger (x) c(x) [\psi^\dagger_R \psi_L e^{i2k_Fx}
+ h.c.] .
\end{equation}
Unfortunately, to leading order this interaction
does not contribute to the low temperature scattering rate.
To see this, consider the scattering process
which transfers $2k_F$ momentum but zero energy to the heavy particle.
Energy and momentum conservation require, $\epsilon_k = \epsilon_{k'}$
and $k-k'=2k_F$, where $k$ and $k'$ are the initial and final
particle momenta.  Together, these imply $k=-k'=k_F$,
which corresponds to a large particle energy, $\epsilon_{k_F} = (m/M)E_F$.
At temperatures below this energy scale, this process
will freeze out.

However, higher order processes which are generated by $H_{int,2k_F}$
will contribute to the low temperature scattering.
Specifically, consider the interaction term,
\begin{equation}
H_{eff} = \tilde{\lambda} \int_x c^\dagger (x) c(x) \psi^\dagger_R \psi_R
\psi^\dagger_L \psi_L   ,
\end{equation}
which will be generated by $H_{int,2k_F}$ at second order.
The coupling constant is
$\tilde{\lambda} = U_0^2/\epsilon_{2k_F}$,
where the denominator, $\epsilon_{2k_F}$, is the energy of the
heavy particle in the
``intermediate state".
This interaction term can be readily bosonized using (2.7)-(2.8),
giving
\begin{equation}
H_{eff} = \lambda \int_x c^\dagger (x) c(x) [(\partial_x \theta)^2
- (\partial_x \phi)^2]  ,
\end{equation}
with $\lambda = \tilde{\lambda}/4\pi$.

The scattering rate from the process $H_{eff}$
can be computed using Fermi's Golden rule
giving,
\begin{equation}
\Gamma_{k \rightarrow k+q} = a_g
[\omega_q^2 - (\Delta \epsilon)^2] n_{(\omega_q+\Delta \epsilon)/2}
[ n_{(\omega_q-\Delta \epsilon)/2} +1]  ,
\end{equation}
with
$a_g = \frac{\lambda^2 }{8Lg^2v^3} (1+ g^4)$ and  $\Delta \epsilon =
\epsilon_{k+q} - \epsilon_q$.
As required, $\Gamma$ satisfies
a detailed balance condition, $f_0(k) \Gamma_{k \to p} = f_0(p) \Gamma_{
p \to k}$, where $f_0(k) = (const)e^{-\beta \epsilon_k}$ is the
equilibrium momentum distribution function for the heavy particle
at temperature $T$.
Notice that this rate has appreciable weight
at small energy and momentum transfer, vanishing as a power rather
than exponentially.  This leads to a power law dependence
of the mobility $\mu(T)$ on temperature, as we now demonstrate.

The mobility can be obtained by solving a
Boltzmann equation for the momentum distribution function, $f(p,t)$,
in the presence of an applied electric field $E$:
\begin{equation}
\partial_t f(p,t) + E \partial_p f(p,t) =
I(p,t)  .
\end{equation}
As usual, the ``collision integral" is expressed in terms of
the scattering rates, (4.13), into and out of the state $p$:
\begin{equation}
I(p) = \sum_k \left[f(k,t) \Gamma_{k \to p} - f(p,t) \Gamma_{p \to k}
\right]  .
\end{equation}
We seek a solution of the form $f(k) = f_0(k) G(k)$,
and determine $G(k)$. The collision term can be re-expressed as,
\begin{equation}
I(p) = \sum_q \Gamma_{p+q \to p} f_0(p+q) \left[G(p+q)-G(p)\right].
\end{equation}
Due to the Bose factors in (4.13), the scattering rate $\Gamma$
is a sharply peaked function of the momentum transfer, $q$,
with width $q \sim v T$.  At low temperatures
it is then legitimate to expand both $f_0(p+q)$ and $G(p+q)$
for small $q$.  Moreover, in the low temperature limit, $\Delta \epsilon$
in (4.13) can be set to zero, and
the scattering rate simplifies:
\begin{equation}
\Gamma_{k \to k+q} \to \Gamma_q \equiv a_g \omega_q^2 n_{\omega_q/2}
[n_{\omega_q/2} + 1]  .
\end{equation}
This requires,
\begin{equation}
\frac{\Delta \epsilon}{T} \sim
(v_Fq/T)(k/Mv_F) \sim
\sqrt{T/Mv_F^2} << 1   ,
\end{equation}
where we have used the fact that $v_Fq/T \sim 1$
and $k \sim \sqrt{MT}$.
In this low temperature regime, the collision integral
can be written,
\begin{equation}
I(p) = A\partial_p f_0 \partial_p G + \frac{1}{2} A f_0 \partial_p^2 G ,
\end{equation}
where we have defined
\begin{equation}
A = \sum_q q^2 \Gamma_q  = (const) T^5  ,
\end{equation}
and used the fact that
$\sum_q q \Gamma_q =0$.

With this form for the collision integral, the steady
state Boltzmann equation reduces to a differential equation for $G(p)$:
\begin{equation}
A \partial_p G - EG = \frac{M}{2 \beta p} (A \partial_p^2 G - 2E
\partial_p G )  .
\end{equation}
We now specialize to the linear response limit, for small
electric fields.  To linear order in $E$, the terms on the right side
can be dropped, and the equation readily integrated
to give, $G(p) = (const) e^{Ep/A}$.
The momentum distribution function, $f=f_0G$,
is then given by
\begin{equation}
f(p) = f_0(p-\frac{EM}{\beta A}) .
\end{equation}
The linear response mobility readily follows ,
\begin{equation}
\mu = \frac{<v>}{E} = \frac{1}{ME} \int_p p f(p) = \frac{1}{\beta A} .
\end{equation}
Since $A \sim T^5$, we deduce a mobility which
diverges as $\mu \sim T^{-4}$.
This result agrees with a strong coupling analysis\cite{nos} based on
the Brownian motion of solitons and calculations of the diffusion
coefficient in real space \cite{kagan}.

\section{Conclusion}

In this paper we have analyzed the dynamics of a heavy
particle moving in a 1d repulsively interacting
Luttinger liquid.
The behavior of the particle's mobility
depends on whether the temperature
is larger or smaller than a crossover scale,
$T^* \sim (m/M)E_F$.
Above $T^*$ the mobility
is roughly independent of temperature and proportional
to the conductance $g$ of the Luttinger liquid.
Below $T^*$ the mobility grows upon cooling,
and diverges in the zero temperature limit as
$\mu(T) \sim T^{-4}$.  {\it At} zero temperature,
the heavy particle moves ballistically, with
a renormalized mass.

We thank A.O. Caldeira, A.O. Gogolin, A.W.W. Ludwig and N.V.Prokof'ev
for many illuminating discussions.
We are also grateful to the National Science Foundation for
support under grants PHY94-07194, DMR-9400142 and DMR-9528578.

\end{document}